\documentclass{article}

\usepackage{PRIMEarxiv}

\usepackage[utf8]{inputenc} 
\usepackage[T1]{fontenc}    
\usepackage{hyperref}       
\usepackage{url}            
\usepackage{booktabs}       
\usepackage{amsfonts}       
\usepackage{nicefrac}       
\usepackage{microtype}      
\usepackage{lipsum}
\usepackage{fancyhdr}       
\usepackage{graphicx}       
\usepackage{subcaption}
\usepackage{graphicx}

\title{Aerodynamic Optimization of the Angle of Attack of the Wing Design of a MQ-9 Reaper UAV}

\author{
  Ryan Kim \\
  \emph{Thomas Jefferson High School for Science and Technology}\\
}

\begin{document}
\maketitle

\begin{abstract}
In this paper I present the aerodynamic optimization of the angle of attack of MQ-9 Reaper wings under turbulent airflow conditions. The background discusses the evolution of UAVs such as the Hunter and their usages for different tasks, specifically within the last decade. The main procedural stage consists of creating a criterion for choosing the most aerodynamic wing design by considering certain aerodynamic properties and analyzing the lift and drag coefficients using computer software. At the end of the paper, the final optimized MQ-9 wing design, along with its respective design parameters, are presented and discussed.
\end{abstract}

\keywords{UAV \and Aerodynamics \and Airfoil}

\section{Introduction}
An unmanned aircraft system (UAS) comprises unmanned aerial vehicles (UAVs), control stations, and aircraft computer systems. UASs can be beneficial as they are remote locations where operators can set up communication systems to control the UAV aircraft and analyze certain data. Of the parts of an UAS, the UAV is worthy of additional consideration because it can be physically optimized to increase aerodynamic efficiency. UAVs are able to transmit data back to the control stations about certain information such as altitude, aircraft speed, or even the detection of a target. Many UAVs are designed to automatically respond to certain conditions, giving them a level of intelligence. For example, an intelligent UAV may return back to the starting destination if it runs out of fuel [1]. It’s important to distinguish between drones, which can only execute a pre-programmed task without making any changes during flight, and UAVs, which send information back to the control stations and execute certain tasks based on the decisions made by the controller.

Benefits of UAVs include a lack of an airborne crew, low operational costs, ease of execution, and the convenience of air travel [2]. Some applications include: photography, military and domestic surveillance, fire detection and control, surveying for mapping, and weather monitoring. Although this study does not aim to design the interior or even the fuselage system of UAVs, it is worth noting that 220 kg of payload mass is removed just by locating the control system at a remote location instead of keeping it on the aircraft. Many industries have started to mass-produce UAV aircraft due to these benefits.

Due to its wide range of uses, the UAV has drastically increased in popularity in the last few decades. The benefits and versatility UAVs offer has led to several countries investing in the rapid development and implementation of better designs, with the majority of the UAV market lying in the military [3]. One type of UAV in particular, the Hunter UAV, is commonly used in warfare for surveillance, espionage, target acquisition, and much more, especially by the US Army.

The motivation of this study is to optimize the airfoil geometry of a specific aircraft model,  the MQ-9, which is a newer and upgraded version of the MQ-1. The “M” refers to the Department of Defense designation of multi-role, and the “Q” means that the aircraft is remotely controlled. The MQ-9 is relevant currently, as tensions rising between Ukraine and Russia have incentivized the US to send MQ-9 and similar UAVs to the aid of Ukraine [4 ukraine]. Optimization was done by changing its angle of attack to maximize its lift-to-drag (LD) ratio, a common measurement of aerodynamic efficiency. Once the optimization process is finished, the airfoil is then applied to a 3D CAD model and compared to the original design to see the aerodynamic improvements.

A plethora of studies examining the aerodynamic design of UAVs already exist. For example, Panagiotou et al. conducted a study on Medium-Altitude-Long-Endurance (MALE) UAVs, emphasizing the conceptual and preliminary design phases, while also running computer simulations as a part of their optimization technique [2]. While their study was not specialized for Hunter UAVs (which are more commonly used for close-range operations compared to the MALE UAVs which are more commonly used for longer distance operations), the methodology used in their optimization study was rigorous and can be applied to the study of Hunter UAVs.
	
In Panagiotou et al. study, the first stage was to create a conceptual design of the MALE UAV [2]. Eight engineers were divided into four groups and they all came up with optimal UAV designs with different fuselages, tails, and wings using CAD software. Many limitations were taken into account, such as the total weight of the aircraft which included the payload weight, fuel weight, and empty weight, temperature, maximum speed, and much more. These limitations reflected the terrain They also used the Breguet equation to estimate the required fuel for the flight of the UAV. With each of the four optimal designs that were created, the thrust-to-weight ratio (T/W) and the wing loading (W/S) were calculated. The findings were that the team 2 had the best overall L/D ratio, with the design of a boom-mounted tail configuration featuring raked-type winglets and an airfoil-shaped fuselage.

Another study performed by Kontogiannis and Ekaterianris focused on small-sized UAVs, again utilizing a general design that could possibly be implemented for different UAVs used for various tasks[5]. Again, they optimized wing geometry after picking an optimal existing fuselage, that being the PARSONS-F2-49. Optimized wing features included: wing twist, airfoil angle of attack, and type of winglet design. The study presented here, in contrast, focuses solely on the Hunter UAV, which is used for military operations. 

	In the proceeding sections, the procedural stages and methodology are thoroughly discussed, the results of the simulation runs are discussed and the most aerodynamic UAV is chosen, and the wing and airfoil are optimized on the best UAV model, and finally the conclusion and final UAV design concept are presented.

\section{Methodology}
In this experiment, the GW-19 airfoil and its angle of attack (AoA) is incrementally changed to find the optimal AoA. In aerodynamics, it is generally agreed upon that the best metric for aerodynamic efficiency is the lift-to-drag (L/D) ratio [6]. This study, therefore, seeks to maximize the L/D ratio in order to determine the most aerodynamic AoA. 

The program of choice was Flowsquare 4.0, which solves high-order differential equations to simulate two-dimensional, incompressible flows [7]. The chosen input airfoil was the GW-19 airfoil as it is the root airfoil of the MQ-9 Reaper. Every angle was created in MATLAB and made into an input boundary condition file for Flowsquare to read and use in the simulation as shown in Figures 1-3 [8]. For the Flowsquare parameter file, which was labeled grid.txt, certain conditions were used to simulate the real environment that the MQ-9 Reaper UAV flies in, assuming clear weather. UAVs such as the MQ-9 Reaper operate at altitudes below 9,100 m and have a broad variety of altitudes to cruise in based on the mission [9]. The altitude of 5,000 m above sea level was chosen for simplicity. The average pressure at an altitude of around 5,000 m is 16.22 in Hg (mercury), which is approximately 54,930 Pa of pressure [10]. At this altitude, air has a density of 0.746 kg/m-3 [11] and, according to the US Standard Atmosphere 1976 model, has a dynamic viscosity of around 3.4 * 10-7 lb*s/ft2, which in standard units is 1.63 * 10-5 kg/(m*s). The last parameter needed for the simulation is the relative velocity between the aircraft and the air, which will stay relatively constant for the majority of the flight. Because drag and lift are associated with the velocity squared, the maximum speed is chosen to clearly show the difference between the different AoAs in their respective L/D ratios. The maximum speed of the MQ-9 Reaper is approximately 240 knots, or 123.467 m/s [12]. 

Other Flowsquare parameters were chosen purely based on simulation time constraints and preference, and every simulation run kept the same parameters. A high-order scheme with a 4th order difference and 3rd order time integral was used for the numerical scheme in the simulation. 5000 was the last time step used and each time step lasted around 6.1138 x 10-5 seconds. The reason for this choice was because when preparing the simulation by running pre-trials, after around 3000 time steps, there seemed to be no changes at all to the x and y velocity vector fields. A simulation was run for each AoA, 0 through 20 degrees in increments of 1 degree. Examples of the boundary condition with the airfoil at various angles are provided in Figures 3-5. It is generally advised to not go above around 17 degrees before airflow above the upper surface becomes detached, a phenomenon known as flow separation [13].

The cruise velocity can be helpful to calculate the aircraft range to find the fuel efficiency of the aircraft. Using the lift-drag analysis provided by Flowsquare, the instantaneous air density and velocities were read from the collected simulation data, and then converted to the ratios of drag and lift forces to drag (CD) and lift coefficients (CL), respectively. Next, the drag and lift coefficients were approximated using NASA FoilSim JS, whose values are experimentally derived [14]. The parameters used for FoilSim were a 2.2\% maximum camber at 44.2\% chord and a maximum thickness of 5.4\% at 20.6\% chord, that match the Drela GW-19 parameters retrieved from the NACA database [15 airfoiltools]. The resulting drag and lift forces were calculated by multiplying the reference area and wing surface area, the produced Flowsquare quantities for pressure and velocity, and the drag and lift coefficients, respectively, as shown in equations (1) and (2).

\begin{equation}
    \frac{D}{C_D} = \frac{\rho V^2 A}{2}
    \end{equation}
    \begin{equation}
    \frac{L}{C_L} = \frac{\rho V^2 A}{2}
\end{equation}

\begin{figure}[h]
  \centering
  \includegraphics[scale = 0.75]{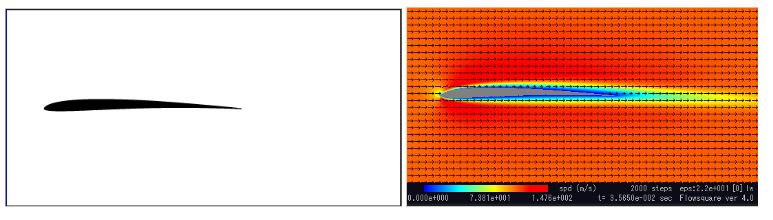}
  \caption{Drela GW-19 airfoil with an AoA of 0 degrees in simulation}

  \includegraphics[scale = 0.75]{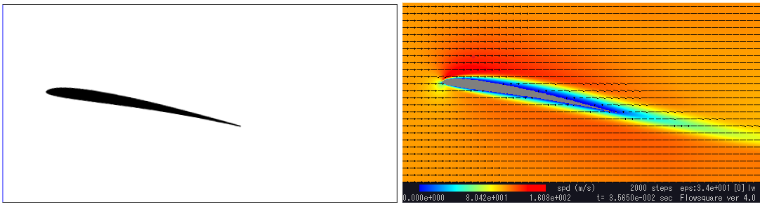}
  \caption{Drela GW-19 airfoil with an AoA of 10 degrees in simulation}

  \includegraphics[scale = 0.75]{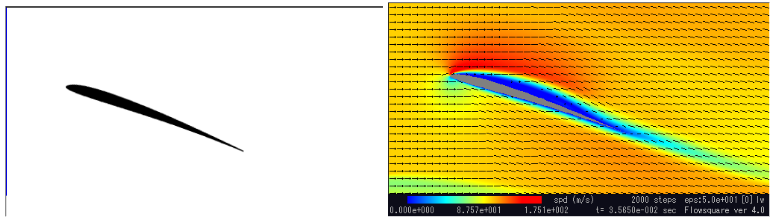}
  \caption{Drela GW-19 airfoil with an AoA of 20 degrees in simulation}
  \label{fig:fig3}
\end{figure}

\newpage

\section{Data Analysis}

\begin{figure}[ht]
  \begin{subfigure}{0.3\textwidth}
    \includegraphics[width=\linewidth]{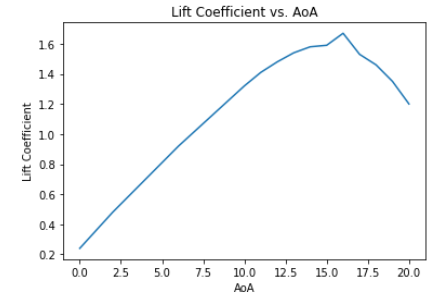}
    \caption{Lift coefficient vs angle of attack} \label{fig:1a}
  \end{subfigure}%
  \hspace*{\fill}   
  \begin{subfigure}{0.3\textwidth}
    \includegraphics[width=\linewidth]{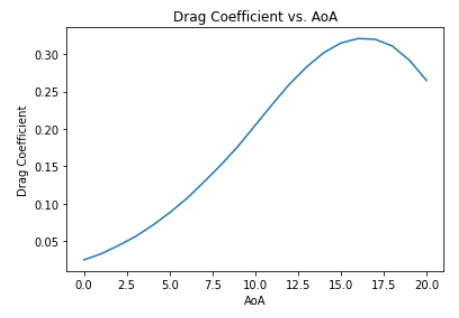}
    \caption{Drag coefficient vs angle of attack} \label{fig:1b}
  \end{subfigure}%
  \hspace*{\fill}   
  \begin{subfigure}{0.3\textwidth}
    \includegraphics[width=\linewidth]{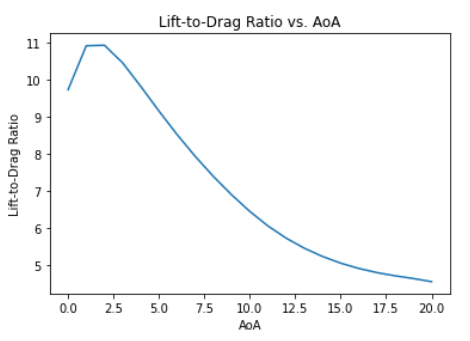}
    \caption{Lift/Drag ratio vs angle of attack} \label{fig:1c}
  \end{subfigure}

\caption{}
\end{figure}

In Figure a, the CL vs AoA graph, the CL steadily increases as the angle of attack increases until the AoA reaches around 16 degrees, when it suddenly begins to decrease. Generally, both the lift and drag coefficients gradually increase as the AoA is increased, until the critical angle of attack. The critical angle of attack occurs when airflow is separated from the top surface of the wing and results in a decrease in the lift coefficient. This period in which rather smooth airflow is converted into turbulent flow is called stall and the drag coefficient can change rapidly and unpredictably. In Figure 6c, the L/D ratio vs AoA graph, the peak L/D ratio is around 2 degrees AoA. This peak is where the wings are able to capture enough airflow on the bottom surface of the wings while capturing minimal airflow that pushes the wings in the opposite direction of flight.

Sharp curves or discontinuities in the graph are a result of the large increment that was used for the AoA. Repeating the experiment using a smaller increment would produce a smoother curve between the integer AoAs. Based on the results of this study, the AoA that produces the highest L/D ratio for a Drela GW-19 airfoil (i.e., the most aerodynamic and fuel-efficient AoA), is approximately 2 degrees.

\section{Conclusion}
The present study attempts to optimize MQ-9 Unmanned-Aerial-Vehicle (UAV) wing design by finding the angle-of-attack (AoA) that will produce the best lift-to-drag (L/D) ratio. The flow conditions were formulated before simulating the wings in a turbulent flow model for each angle of attack. Specific parameters such as the air density, dynamic viscosity, and speed were considered and placed into the grid.txt file which the Flowsquare software utilizes. A Reynolds-Averaged-Navier-Stokes (RANS) model was used in Flowsquare due to eddies not being an integral part of calculating lift and drag forces and certain time constraints. After each simulation for each AoA was completed, the velocity and pressure data files produced by Flowsquare were converted into quantities that represented drag to drag coefficient ratios and lift to lift coefficient ratios through C code. To calculate the lift and drag coefficients, the NASA FoilSim Js software was used to approximate these missing quantities by using the camber height, thickness, and AoA for each trial. The resulting lift and drag forces were calculated and the lift-to-drag ratios were graphed with the AoAs.

The results of the study suggest that the best AoA for the Drela GW-19 airfoil used in the MQ-9 Reaper is approximately 2 degrees. Future work would include implementing this specific airfoil in a CAD model of an MQ-9 Reaper and using a 3D CFD software such as ANSYS Fluent or Autodesk CFD to simulate a wind tunnel similar to the one created in this study. Some other areas of future work would include smaller increments of AoAs for better precision and experimental verification of results using a UAV model in a wind tunnel. I have started to test a 3D CAD Model of the proposed design, as shown in Figures 5-6.

\begin{figure}[ht]
    \includegraphics[width=\linewidth]{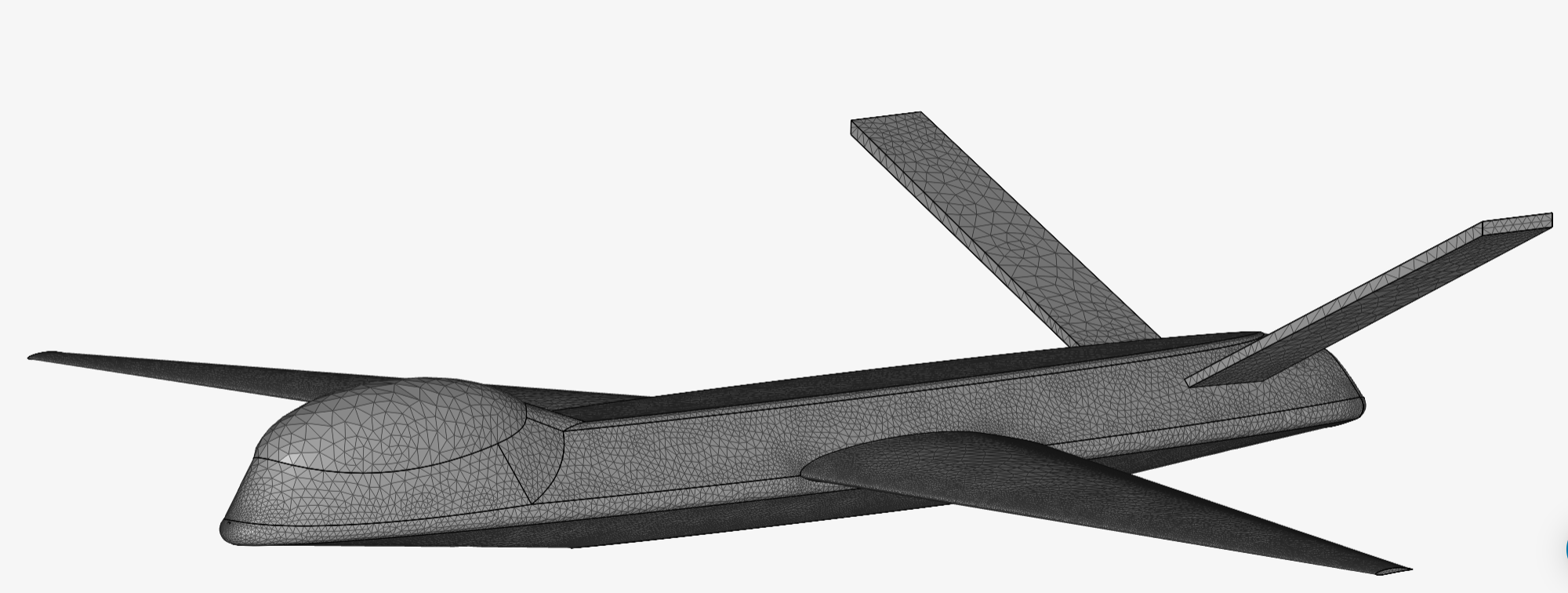}
    \caption{A mesh layout on the proposed MQ-9 Reaper design} \label{fig:2}
\end{figure}%

\begin{figure}[ht]
    \includegraphics[width=\linewidth]{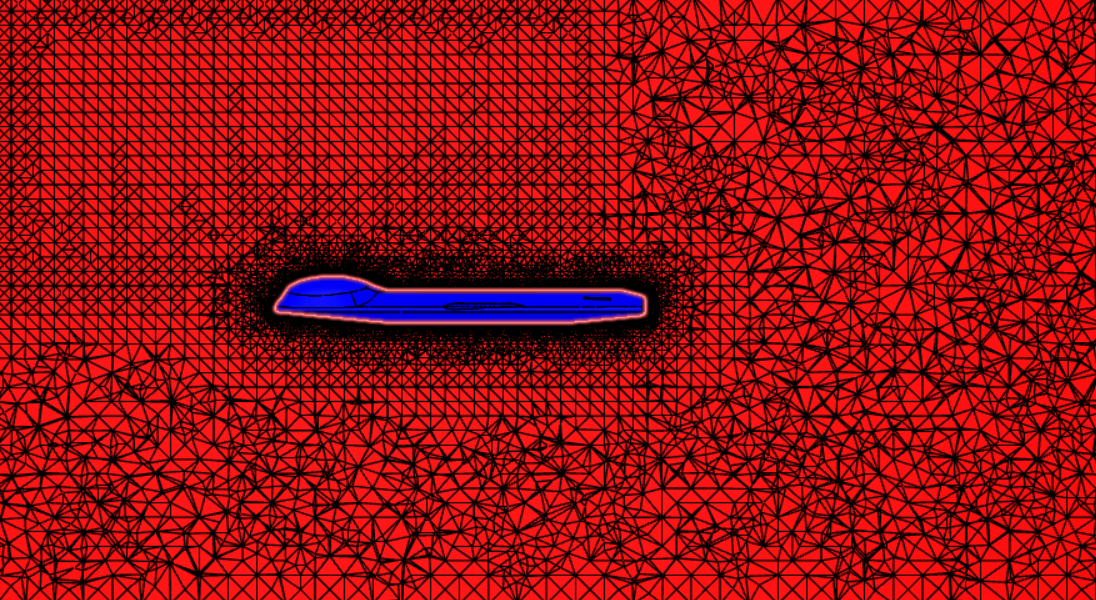}
    \centering
    \caption{The cross section of a simulation run using the proposed MQ-9 Reaper design} \label{fig:3}
\end{figure}

\newpage
\section*{Acknowledgments}
This work made use of Flowsquare, the free, integrated two-dimensional computational fluid dynamics software (http://flowsquare.com). I would also like to thank Haley Wohlever, PhD student at the University of California, Berkeley, for introducing me to and teaching me the basics of fluid dynamics, and for mentoring me throughout this project.
\newpage
\bibliographystyle{unsrt}  
\bibliography{references}
\raggedbottom

\linespread{1.6}
[1] Austin, R. (2010). Unmanned aircraft systems: UAVS design, development and deployment. Wiley.

[2] Panagiotou, P., Kaparos, P., Salpingidou, C., \& Yakinthos, K. (2016). Aerodynamic design of a MALE UAV. Aerospace Science and Technology, 50, 137-138. https://doi.org/10.1016/j.ast.2015.12.033 

[3] Nex, F., Armenakis, C., Cramer, M., Cucci, D.A., Gerke, M., Honkavaara, E., Kukko, A., Persello, C., \& Skaloud, J. (2022). UAV in the advent of the twenties: Where we stand and what is next. ISPRS Journal of Photogrammetry and Remote Sensing, 184, 215-242. https://doi.org/10.1016/j.isprsjprs.2021.12.006 

[4] Sirota, S. (n.d.). Military-industrial complex is itching to send "hunter-killer" drones to Ukraine. The Intercept. https://theintercept.com/2022/05/18/ukraine-reaper-drones-weapons-transfer/ 

[5] Kontogiannis, S., \& Ekaterinaris, J. (2013). Design, performance evaluation and optimization of a UAV. Aerospace Science and Technology, 29(1), 339-350. https://doi.org/10.1016/j.ast.2013.04.005 

[6] Lift to Drag Ratio. (n.d.). NASA Glenn Research Center. Retrieved September 2, 2022, from https://www.grc.nasa.gov/www/k-12/airplane/ldrat.html 

[7] Flowsquare (Version 4.0) [Computer program].

[8] MATLAB. (n.d.). https://www.mathworks.com/products/matlab.html 

[9] Chamola, V., Kotesh, P., Agarwal, A., Naren, Gupta, N., \& Guizani, M. (2021). A comprehensive review of unmanned aerial vehicle attacks and neutralization techniques. Ad Hoc Networks, 111. https://doi.org/10.1016/j.adhoc.2020.102324

[10] Altitude pressure chart. (n.d.). ITT Aerospace.

[11] Intercept. https://theintercept.com/2022/05/18/ukraine-reaper-drones-weapons-transfer/ 
U.S. Standard Atmosphere vs. Altitude. (n.d.). The Engineering Toolbox. Retrieved September 2, 2022, from https://www.engineeringtoolbox.com/standard-atmosphere-d\_604.html

[12] MQ-9 Reaper. (n.d.). US Air Force. Retrieved September 2, 2022, from https://www.af.mil/About-Us/Fact-Sheets/Display/Article/104470/mq-9-reaper/

[13] Angle of attack awareness [Infographic]. (n.d.). FAA Aviation Safety. https://www.faa.gov/news/safety\_briefing/2019/media/SE\_Topic\_19\_04.pdf

[14]FoilSim Student JS [Computer software]. (n.d.). https://www.grc.nasa.gov/www/k-12/airplane/foil3.html 

\end{document}